\documentstyle[prl,aps]{revtex}

\begin{document}
\draft
\twocolumn

\input epsf
\title{ Self-control in Sparsely Coded Networks }

\author{D.R.C. Dominguez \cite{email1} and D. Boll\a'e \cite{email2}}
\address{Instituut voor Theoretische Fysica,
Katholieke Universiteit Leuven, B-3001 Leuven, Belgium}

\date{\today}
\maketitle


\begin{abstract}
A {\it complete self-control} mechanism is proposed in the dynamics of
neural networks through the introduction of a time-dependent threshold,
determined in function of both the noise and the pattern activity in the
network.
Especially for sparsely coded models this mechanism is shown to
considerably improve the storage capacity, the basins of attraction and
the mutual information content of the network.
\end{abstract}


Sparsely coded models have attracted a lot of attention in the development
of neural networks, both from the device oriented and biologically
oriented point of view  \cite{NT90}--\cite{Ok96}. It is well-known that
they have a large storage capacity, which behaves as $1/(a \ln a)$ for $a$
small where $a$ is the pattern activity. However, it is clear that the
basins of attraction, e.g., should not become too small because then
sparse coding is, in fact, useless.

In this context the necessity of an activity control system has been
emphasized, which tries to keep the activity of the network in the
retrieval process the same as the one for the memorized patterns
\cite{AGS87}--\cite{BDS89}. This has led to several
discussions imposing external constraints on the dynamics
(see the references in \cite{Ok96}). Clearly, the
enforcement of such a constraint at every time step destroys part of the
autonomous functioning of the network.

An important question is then whether the capacity of storage and retrieval
with non-negligible basins of attraction can be improved and even be
optimized without imposing these external constraints, keeping at the same
time the simplicity of the architecture of the network.

In this Letter we answer this question by proposing, as far as we are
aware for the first time, a {\it complete self-control}
mechanism in the dynamics of neural networks. This is done through the
introduction of a time-dependent threshold in the transfer function. This
threshold is chosen as a function of the noise in the system and the
pattern activity, and adapts itself in the course of the time evolution.
The difference with existing results in the literature \cite{Ok96}
precisely lies in this adaptivity property. This immediately solves, e.g.,
the difficult problem of finding the mostly narrow interval for an
optimal threshold such that the basins of attraction of the memorized
patterns do not shrink to zero.

We have worked out the practical case of sparsely coded models. We find
that the storage capacity, the basins of attraction as well as the
mutual information content are improved. These results are shown to be
valid also for not so sparse models. Indeed, a similar
self-control mechanism should even work in more complicated architectures,
e.g., layered and fully connected ones. Furthermore, this idea of
self-control might be relevant for dynamical systems in general, when
trying to improve the basins of attraction and the convergence times.

\vspace{-4cm}
\begin{figure}[t]
\begin{center}
\epsfysize=10cm
\leavevmode
\epsfbox[1 1 600 800]{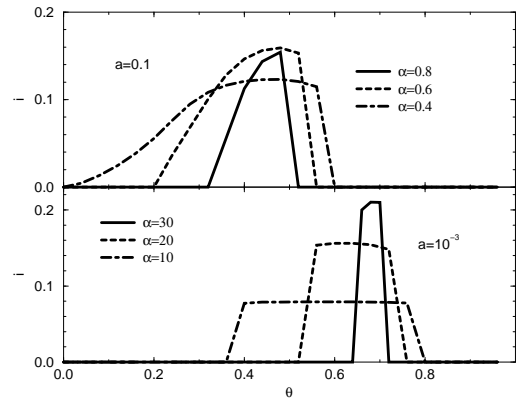}
\end{center}
\caption{ The information $i$ as a function of $\theta$ without
self-control for $a=0.1$ (top) and $a=0.001$ (bottom) for several values
of $\alpha$.}
\end{figure}

\vspace{-4cm}
\begin{figure}[t]
\begin{center}
\epsfysize=10cm
\leavevmode
\epsfbox[1 1 600 800]{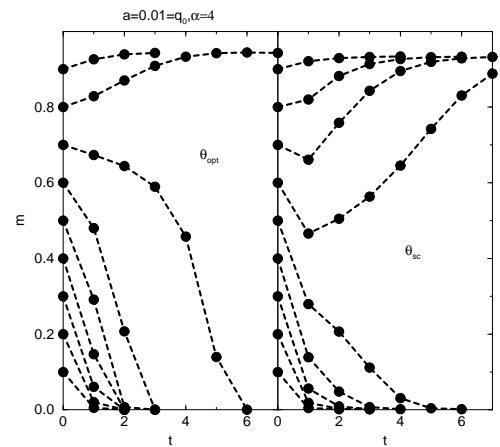}
\end{center}
\caption{ The evolution of the overlap $m_{t}$ for several initial
values $m_{0}$, with $q_{0}=0.01=a$ and $\alpha=4$ for the
self-control model (right) and the optimal threshold model (left). The
dashed curves are a guide to the eye.}
\end{figure}

Consider a network of $N$ binary neurons. At time $t$ and zero temperature
the neurons $\{\sigma_{i,t} \} \in \{0,1\}, \,\,\, i=1, \ldots, N$ are
updated in parallel according to the rule
\begin{equation}
     \sigma_{i,t+1}= F_{\theta_{t}}(h_{i,t}), \,\,\,
          h_{i,t}= \sum_{j(\neq i)}^{N}J_{ij}(\sigma_{j,t}-a)\, .
      \label{2.si}
\end{equation}
In general, the input-output relation $F_{\theta_{t}}$ can be a monotonic
function with $\theta_{t}$ a time-dependent threshold.
In the sequel we restrict ourselves to the step function
$F_{\theta_{t}}(x)= \Theta(x-\theta_{t})$.
The quantity $h_{i,t}$ is the local field of neuron $i$ at time $t$ and
$a$ is the activity of the stored patterns, $\xi^{\mu}_{i} \in\{0,1\}
\,\,\, \mu=1, \ldots, p$. The latter are independent identically
distributed random variables (IIDRV) with respect to $i$ and $\mu$
determined by the probability distribution
\begin{equation}
   p(\xi^{\mu}_{i})= a\delta(\xi^{\mu}_{i}-1)
                                 +(1-a)\delta(\xi^{\mu}_{i}).
        \label{2.px}
\end{equation}
At this point we remark that the activity can be written as $a=(1-b)/2$
with $-1<b<1$ the bias of the patterns as defined, e.g., in \cite{AGS87}.
In fact, $\langle\xi^{\mu}_{i}\rangle=a$ but no correlations between
the patterns occur, i.e., $\langle\xi^{\mu}_{i}\xi^{\nu}_{i}\rangle
-\langle\xi^{\mu}_{i}\rangle\langle\xi^{\nu}_{i}\rangle=0$.
We now consider an extremely diluted asymmetric version of this
model in which each neuron is connected, on average, with $C$ other
neurons. In that case the synaptic couplings $J_{ij}$ are determined
by the covariance rule
\begin{equation}
    J_{ij}= {C_{ij}\over C{\tilde a}}
           \sum_{\mu=1}^{p} (\xi^{\mu}_{i}-a)(\xi^{\mu}_{j}-a),
           \quad   {\tilde a}\equiv a(1-a)\,.
    \label{2.Ji}
\end{equation}
Here the $C_{ij}\in\{0,1\}$ are IIDRV with probability
$Pr\{C_{ij}=1\}=C/N \ll 1, C>0$. For $a=1/2$ and $\theta_t=0$ we recover
the diluted Hopfield model.

The relevant order parameters measuring the quality of retrieval are the
overlap of the microscopic state of the network and the $\mu$th
pattern, and the neural activity
\begin{equation}
   m^{\mu}_{N,t}\equiv \frac{1}{Na}\sum_{i}\xi^{\mu}_{i}\sigma_{i,t},
   \quad   q_{N,t}\equiv {1\over N}\sum_{i}\sigma_{i,t}\,.
    \label{2.Mm}
\end{equation}
The $m^{\mu}_{N,t}$ are normalized order parameters within the interval
$[-1,1]$, which attain the maximal value $m^{\mu}_{N,t}=1$. They have to
be considered over the diluted structure such that the loading $\alpha $
is defined by $p=\alpha C$. The Hamming distance between the state of the
neuron and the pattern $\xi^{\mu}$ can be written as
$ d^{\mu}_{t} = a- 2am^{\mu}_{N,t}+q_{N,t}.$

To fix the ideas and without loss of generality, we take an initial
network configuration correlated with only one pattern meaning that only
the retrieval overlap for that pattern, say $\mu=1$, is macroscopic, i.e.,
of order ${\cal O}(1)$ in the thermodynamic limit $C,N \rightarrow \infty$.
The rest of the patterns causes a residual noise at
each time step of the dynamics. Depending on the architecture of the
network this noise might be extremely difficult to treat \cite{BKS90}.
A novel idea is then to let the network itself autonomously counter this
residual noise at each step of the dynamical evolution, by introducing
an adaptive, hence time-dependent, threshold. We propose the general form
$\theta_{t}(x)= c(a) [\mbox{Var}(\omega_{t})]^{1/2}$ with $\omega_{t}$ this
residual noise. This self-control mechanism of the network is complete if
we find a way to determine $c(a)$.

\vspace{-4cm}
\begin{figure}[t]
\begin{center}
\epsfysize=10cm
\leavevmode
\epsfbox[1 1 600 800]{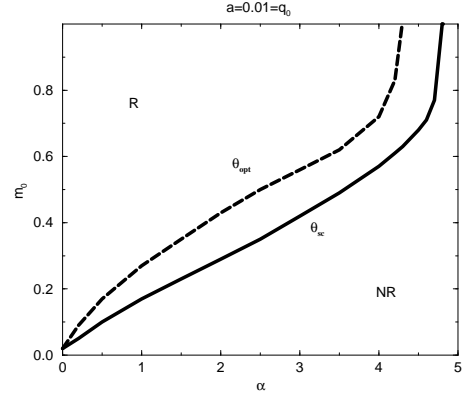}
\end{center}
\caption{ The basin of attraction as a function of $\alpha$ for
$a=0.01$ and initial $q_{0}=a$ for the self-controlled model
(full line) and the optimal threshold model (dashed line). }
\end{figure}

\vspace{-5cm}
\begin{figure}[t]
\begin{center}
\epsfysize=10cm
\leavevmode
\epsfbox[1 1 600 800]{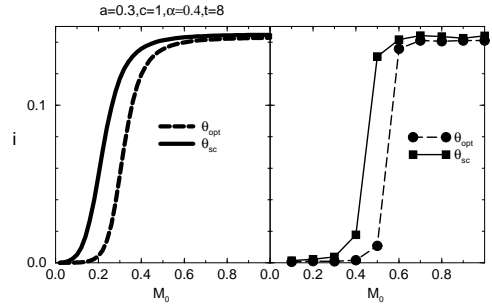}
\end{center}
\caption{ The information $i$ as a function of $M_0$ for $a=0.3,
\alpha=0.4$ with ($c=1$) and without self-control. Left: analytic
results. Right: simulations up to $t=8$ for $N=47.000, C=50, p=20$
averaged over $10$ samples. }
\end{figure}

In order to do so we first write down the evolution equations governing the
dynamics. We recall that for the particular model we are considering the
parallel dynamics can be solved exactly following the methods involving a
signal-to-noise analysis (see, e.g, \cite{DGZ87}, \cite{BSVZ94}). Such an
approach leads to the following equations for the order parameters in
the thermodynamic limit $C,N \rightarrow \infty$
\begin{eqnarray}
  &&m_{t+1}^1=
     \langle F_{\theta_{t}}[(1-a)M^1_{t}+\omega_{t}] \rangle_{\omega}
     \label{3.M1}\\
  &&q_{t+1}=a m_{t+1}^1 +
         (1-a)\langle F_{\theta_{t}}(-aM^1_{t}
                       +\omega_{t}) \rangle_{\omega}\, ,
              \label{3.Q}
\end{eqnarray}
with $ M^1_{t}=(m^1_{t}-q_{t})/(1-a)$, where we have averaged over the
first  pattern $\xi^1$ and where the angular brackets indicate that we
still have to average over the residual noise $\omega_{t}$
which can be written as $\omega_{t}=[\alpha Q_{t}]^{1/2} {\cal N}(0,1)$
with $Q_{t}=(1-2a)q_{t} + a^2$ and ${\cal N}(0,1)$ a Gaussian random
variable with mean zero and variance unity. The order parameters
$m^1_t$ and $q_t$ are the thermodynamic limits of (\ref{2.Mm}).
The quantity $M^1_t$ reduces to the overlap of the Hopfield model, again
when taking $a=1/2$ and $\theta_t=0$. From now on we forget about the
superscript $1$.

The equations (\ref{3.M1}) and (\ref{3.Q}) give a self-controlled dynamics
if we can completely specify, a priori, the threshold $\theta_{t}$
proposed before. We remark that, for the present model, $\theta_{t}$
is a macroscopic parameter, thus no average must be done over
the microscopic random variables at each time step $t$. We have, therefore,
a mapping with a threshold with changes each time step but no statistical
history effects the evolution process. What is left then is to find
an optimal form for $c(a)$.

A very intuitive reasoning based on the detailed behavior of these
equations (\ref{3.M1})-(\ref{3.Q}) goes as follows. To have $m \sim 1-
\mbox{erfc}(n)$ and $q\sim a + \mbox{erfc}(n)$ with $n>0$ at a given time
$t$ such that good retrieval properties, i.e., $\sigma_i=\xi_i$  for most
$i$ are realized, we want the following inequalities to be satisfied:
$(1-a)M_t-n[\alpha Q_t]^{1/2}\geq\theta_t$ and $-aM_t+n [\alpha Q_t]^{1/2}
\leq \theta_t$.
Using the general form for the threshold $\theta_{t}$ we obtain that
$2n \sim M_t[\alpha Q_t]^{-1/2}$. This leads to  $c(a) \sim n(1-2a)$. Here
we remark that $n$ itself depends on $\alpha$ in the sense that for
increasing $\alpha$ it gets more difficult to have good retrieval such
that $n$ decreases. But it can still be chosen a priori.

In the limit of sparse coding meaning that the fraction of active neurons
is very small and tends to zero in the thermodynamic limit, we can present
a more refined result for $c(a)$ by rewriting the second term on the r.h.s.
of Eq.~(\ref{3.Q}) asymptotically as
\begin{eqnarray}
   \langle [F_{\theta}[-aM +\omega] \rangle_{\omega}
    = {1\over 2}[1-\mbox{erf}(\frac{aM}{\sqrt{2\alpha Q}}
               +\frac{c(a)}{\sqrt{2}})]
      \to   \frac{e^{-c^{2}/2}}{c \sqrt{2\pi}}
     \label{4s.pJ}
     \nonumber
\end{eqnarray}
This term must vanish faster than $a$ so that we obtain
$c=[-2\ln(a)]^{1/2}$. Using this and the first inequality written down
above we can evaluate the maximal capacity for which some small errors in
the retrieval are allowed. The result is
$ \alpha={\cal O}(|a\ln(a)|^{-1}) $,
which is of the same order as the critical capacity found for
non-self-controlled sparsely coded neural networks \cite{Ok96},\cite{BDS89},
\cite{Ts88}-\cite{Ho89}.

Next, it is known that while the Hamming distance is a good measure for the
performance of a uniform network (i.e., $a\sim1/2$), it does not give a
complete description of the information content for sparsely coded
networks. In more detail, it can not distinguish between a situation where
most of the wrong neurons ($\sigma_i \neq \xi_i$) are turned off and a
situation where these wrong neurons are turned on. This distinction is
extremely critical because the inactive neurons carry less information
than the active ones.
To give one example, when $\sigma_{i}=0$ for all $i$, the Hamming
distance $d=a$ and hence vanishes in the sparsely coded limit, while for
$\sigma_{i}=1$ for all $i$, $d=1-a$ and hence goes to $1$. However, in
both cases there is no information transmitted. To solve
this problem we introduce the mutual information content of the network.

\vspace{-4cm}
\begin{figure}[t]
\begin{center}
\epsfysize=10cm
\leavevmode
\epsfbox[1 1 600 800]{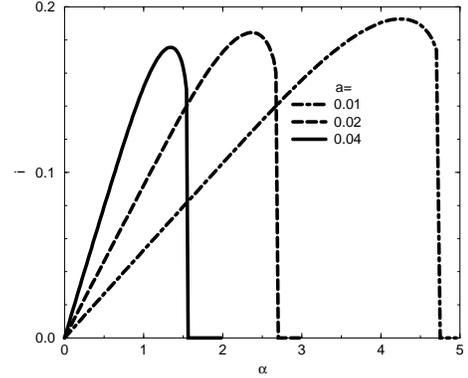}
\end{center}
\caption{ The information $i$ as a function of $\alpha$ for the
self-controlled model with several values of $a$. }
\end{figure}

\vspace{-4cm}
\begin{figure}[t]
\begin{center}
\epsfysize=10cm
\leavevmode
\epsfbox[1 1 600 800]{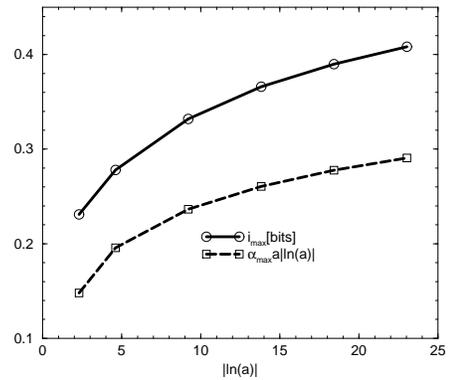}
\end{center}
\caption{ The maximal information $i_{max}/\ln(2)$ and
$\alpha_{max} a|\ln(a)|$ as a function of $|\ln(a)|$. }
\end{figure}

The mutual information function (see, e.g., \cite{B90}) is a concept in
information theory which measures the average amount of information that
can be received by the user by observing the signal at the output of a
channel. For the problem at hand, i.e. retrieval dynamics of the pattern
$\mu=1$, where each time step is regarded as a channel it can be defined
as (we forget about the time index $t$)
\begin{eqnarray}
   && I(\sigma_i;\xi_i)=S(\sigma_i)-
                 \langle S(\sigma_i|\xi_i)\rangle_{\xi_i},
          \label{3.II}\\
   && S(\sigma_i)\equiv -\sum_{\sigma_i}p(\sigma_i)\ln[p(\sigma_i)],
          \label{3.III} \\
   && S(\sigma_i|\xi_i)\equiv
             -\sum_{\sigma_i}p(\sigma_i|\xi_i)\ln[p(\sigma_i|\xi_i)].
          \label{3.Is}
\end{eqnarray}
Here $S(\sigma_i)$ and $S(\sigma_i|\xi_i)$ are the entropy
and the conditional entropy of the output, respectively.
The quantity $p(\sigma_i|\xi_i)$ is the conditional probability that the
$ith$ neuron is in a state $\sigma_{i}$ at time $t$, given that the $ith$
site of the pattern being retrieved is $\xi_{i}$. It is given by
\begin{eqnarray}
p(\sigma|\xi)&=&
     [\gamma_{0}+(m-\gamma_{0})\xi]\delta_{\sigma-1}+
     [1-\gamma_{0}-(m-\gamma_{0})\xi]\delta_{\sigma} ,
     \nonumber\\
     \gamma_{0}&=&{q-am\over 1-a}.
\label{3.ps}
\end{eqnarray}
where we have assumed that this formula holds for every site index $i$,
and where the $m$ and $q$ are precisely the order parameters (\ref{2.Mm})
in the thermodynamic limit. We have also used the normalizations
$\sum_{\sigma}p(\sigma|1)=\sum_{\sigma}p(\sigma|0)=1$.
Using the probability distribution of the patterns (Eq.(\ref{2.px})), we
furthermore obtain
\begin{equation}
    p(\sigma)\equiv\sum_{\xi}p(\xi)p(\sigma|\xi)=
     q\delta(\sigma-1)+(1-q)\delta(\sigma).
   \label{3.px}
\end{equation}
Hence the expressions for the entropies defined above become
\begin{eqnarray}
   S(\sigma)&=& -q\ln q - (1-q)\ln(1-q),\,\,
         \\
    \langle S(\sigma|\xi)\rangle_{\xi}&=&
    -a[m\ln(m)+ (1-m)\ln(1-m)]-
        \nonumber\\
    &&(1-a)[ \gamma_0 \ln \gamma_0 + (1-\gamma_0)\ln(1-\gamma_0)].
    \label{3.Hs}
\end{eqnarray}
Recalling eq.~(\ref{3.II}) this completes the calculation of the mutual
information content of the present model.

We have solved this self-controlled dynamics for the sparsely coded network
numerically and compared its retrieval properties with non-self-controlled
models. We are only interested in the retrieval solutions leading to
$M>0$ and carrying a non-zero information $I$.

In Fig.~1 we have plotted the information content
$i\equiv pNI/\#J =\alpha I$ as a
function of the threshold $\theta$ for $a=0.1$ and $a=0.001$ and different
values of $\alpha$, $without$ self-control. This illustrates that it is
rather difficult, especially for sparse coding, to choose a threshold
interval such that $i$ is non-zero.

In Fig.~2 we compare the time evolution of the retrieval overlap, $m_{t}$,
starting from several initial values, $m_{0}$, for the self-control model
with an initial neural activity $q_{0}=0.01=a$ and $\theta_{sc}=[-2(\ln
a)\alpha Q_t]^{1/2}$, with the model where the threshold
is chosen by hand in an optimal way in the sense that we took the one with
the greatest information content $i$, by looking at the corresponding
results of Fig.~1 for $a=0.01$. We see that the self-control forces more of
the overlap trajectories to go to the retrieval attractor. It does improve
substantially the basin of attraction. This is further illustrated in
Fig.~3 where the basin of attraction for the whole retrieval phase $R$ is
shown for the model with a $\theta_{opt}$ selected for every
loading $\alpha$ and the model with self-control $\theta_{sc}$.
We remark that even near the border of critical storage
the results are still improved. Hence the storage
capacity itself is also larger.
These results are not strongly dependent upon the initial value of $q_0$ as
long as $q_{0}={\cal O}(a)$.

Furthermore, we find that self-control gives a comparable improvement for
not so sparse models, e.g., $a \sim 0.3$. This is illustrated in Fig.~4
where we show some analytic results together with a first set of
simulations for the basins of attraction. This type of simulations for
extremely diluted models is known to be difficult because of the
theoretical limits $C,N \rightarrow \infty$ and $\ln C \ll \ln N$.
Nevertheless, it is clear that concerning the self-control aspect,
qualitative agreement with the analytic results is obtained.
For these values of $a$, $M_0$ is the relevant quantity. The quantitative
difference is mostly due to the fact that $M_0^{anal} \sim
M_0^{simul} + {\cal O}(1/\sqrt{Ca})$.

Figure 5 displays the information $i$ as a function of $\alpha$
for the self-controlled model with several values of $a$.
We observe that $i_{max} \equiv i(\alpha_{max})$ is reached
somewhat before the critical capacity and that it slowly increases with
increasing $\alpha $.

Finally, in Fig.~6 we have plotted $i_{max}/\ln(2)$ and
$\alpha_{max}a|\ln(a)|$
as a function of the activity on a logarithmic scale.
It shows that $i_{max}$ increases with $|\ln(a)|$ until it starts
to saturate. The saturation is rather slow, in agreement with results found
in the literature \cite{Pe89}, \cite{Ho89}.

In conclusion, we have found a novel way to let a diluted network
autonomously control its dynamics such that the basins of attraction and
the mutual information content are maximal.

We thank S.Amari and G.Jongen for useful discussions. This work has been
supported by the Research Fund of the K.U.Leuven (grant OT/94/9). One of
us (D.B.) is indebted to the Fund for Scientific Research - Flanders
(Belgium) for financial support.

\end{document}